\documentstyle[proceedings]{crckapbk}

\begin{opening}
\title{RORF -- a Radio Optical Reference Frame}
\author{N.ZACHARIAS, A.L.FEY, J.L.RUSSELL, K.J.JOHNSTON}
\institute{USRA / NRL / ARC / USNO, Washington D.C.}
\end{opening}

\begin{document}

The radio observations are based on
more than one million pairs of group delay and phase delay rate
observations from {\it all} applicable dual frequency Mark--III VLBI data
from 1979 until the end of 1993.

A subset of 436 sources,
nearly uniformly distributed over the entire sky,
has been selected to define a celestial inertial frame
which is presented to the IAU Working Group on the Radio/Optical
Reference Frame.

The precision of most sources is below 1 mas per coordinate.
A comparison to the JPL94R01 catalog has been made.
The $\Delta \delta$ vs. $\delta$ plot shows an offset of up to
1 mas for southern declinations, while a comparison with the IERSC01
catalog shows no systematic differences larger than 0.5 mas.
Details are given in Johnston et al.(1994).

Because of the faintness of the optical counterparts (V $\approx$ 17...21) of
the
compact extragalactic radio sources a multi--step approach
is required to obtain optical positions with respect to the
optical reference system based on bright stars (FK5, IRS, Hipparcos).

The systematic errors as function of magnitude are controled by use
of objective gratings with the astrographs and different exposure
times with the other instruments, including wide field CCD observations.
Depending on the available data a standard error of 10 to 50 mas can
be obtained for the internal precision of the mean position of a source.
About 95\% of the astrograph work and 30\% of the
source observations are completed.
A precision of about 1 mas for the radio--optical link is feasable
with this technique. A precision of about 2.5 mas is estimated
with currently available data.

The radio catalog is available
via anonymous ftp at maia.usno.navy.mil (192.5.41.22) or via
Mosaic with the URL ``file://maia.usno.navy.mil/rorf''.
The complete poster paper is available from nz@pyxis.usno.navy.mil.

\vspace*{1mm}
\begin{flushleft}
Johnston,K.J., et al. (1994) "A Radio Reference Frame", submitted to {\it AJ}
\end{flushleft}

\end{document}